# Unexpectedly high salt accumulation inside carbon nanotubes soaked in very dilute salt solutions

Xueliang Wang[1,3†], Guosheng Shi[2,4†], Shanshan Liang[2], Jian Liu[5], Deyuan Li[2,4], Gang Fang[2,3], Renduo Liu[1], Long Yan[1*] and Haiping Fang[2*]

[1]Center for Thorium Molten Salt Reactor System, Shanghai Institute of Applied Physics, Chinese Academy of Sciences, Shanghai 201800, China.
[2]Division of Interfacial Water, Key Laboratory of Interfacial Physics and Technology and Shanghai Synchrotron Radiation Facility, Shanghai Institute of Applied Physics, Chinese Academy of Sciences, Shanghai 201800, China.
[3]University of Chinese Academy of Sciences, Beijing 100049, China.
[4]Shanghai Applied Radiation Institute and Department of Physics, Shanghai University, Shanghai 200444, China.
[5]School of Physical Science and Technology, ShanghaiTech University, Shanghai 201210, China

[†] These authors contributed equally to this work
*Corresponding author. E-mail: fanghaiping@sinap.ac.cn (H.-P. F.); yanlong@sinap.ac.cn (L. Y.)

We experimentally demonstrate the formation of salt aggregations with unexpectedly high concentration inside multi-walled carbon nanotubes (CNTs) soaked only in dilute salt solutions and even in solutions containing only traces of salts. This finding suggests the blocking of fluid across CNTs by the salt aggregations when CNTs are soaked in a dilute salt solution with the concentration of seawater or even lower, which may open new avenues for the development of novel CNT-based desalination techniques. The high salt accumulation of CNTs also provides a new CNT-based strategy for the collection/extraction of noble metal salts in solutions containing traces of noble metal salts. Theoretical analyses reveal that this high salt accumulation inside CNTs can be mainly attributed to the strong hydrated cation–π interactions of hydrated cations and π electrons in the aromatic rings of CNTs.

It has been found that pure water can flow exceptionally fast through carbon nanotubes (CNTs) [1-5], which thus shows great potential for use in the design of desalination membranes [6-8], devices for nanofluidic [9-16] and ion separation [6,7] as well as for understanding biological activities [13,17]. In practice, most of these systems for the applications use salt solutions rather than pure water, and the water permeation is often significantly affected by the ions. There have been many theoretical studies [10,18-20] predicting effective ion rejection by CNTs of various diameters simply on the basis of classic molecular dynamics (MD) simulations without carefully considering the ion effects, nevertheless, there is *no direct experimental studies have demonstrated salt rejection adequate for desalination using CNTs* till 2016 [7], even though the fabrication of CNTs has greatly improved [21-25]. Very recently, Tunuguntla et al. [26] reported fast water flow across narrow CNTs (inner diameter ~0.8 nm) in salt solution. In their work, the CNTs were functionalized with saturated carboxylate groups (–COOH) at the ends soaked in salt solution, which is consistent with the previous theoretical prediction attributing the fast water flow to the functionalization at the tube ends preventing cation adsorption on the aromatic rings at the ends of the CNTs [8]. However, for wide CNTs (inner diameter > 1 nm), several experiments [4,23,26] have shown that without the help of pressure or an electric field above a certain threshold, only very little water flow and ion permeation can be achieved across the CNTs. This finding contrasts with the conventional intuition that faster water permeation should occur across wider tubes with the same internal surfaces [27,28]. Clearly, different from the case for the narrow CNTs, ions can enter the wide CNTs; therefore, the effect of the ions inside the wide CNTs on the water and ion permeation becomes important. Unfortunately, there is very little knowledge on the behaviors of ions inside the CNTs. To be worse, the behaviors of the salt solutions inside CNTs soaked in dilute solutions under ambient conditions have never been reported experimentally although there are previous theoretical studies predicting the trapping and adsorption of a single cation and anion inside wide CNTs immersed in aqueous solution [8,29].

Here, we show experimentally that when soaking multi-walled CNTs (inner diameter of 5–15 nm) in dilute salt solutions, and even in solutions containing traces of salts, salt aggregations with unexpectedly high concentration are formed inside the CNTs. These findings indicate the high salt accumulation of CNTs, which is unexpected because from the conventional view, the concentrations of the salt solution inside and outside the CNTs are regarded as being comparable [30-33]. Moreover, the high concentration of the salt aggregations suggests the blocking of the fluid across the CNTs. This discovery helps explain the recent findings of negligible water flow and ion permeation across wider CNTs in salt solution and opens new avenues for the development of novel ideas related to CNT-based desalination. It also provides a new strategy for the collection of noble metal salts in solutions containing traces of noble metal salts, which has great economic value [34] and significance for environmental conservation. Our theoretical studies reveal that this high salt accumulation of CNTs can be mainly attributed to the strong hydrated cation–π interactions between hydrated cations and π electrons in the aromatic rings of CNTs. Furthermore, the strong cation–π interactions between other cations (including $Li^+$, $Mg^{2+}$, $Ca^{2+}$, $Cu^{2+}$, $Cd^{2+}$, $Cr^{2+}$, $Pb^{2+}$, $Ag^+$, $Pd^{2+}$, and $Rh^{3+}$) and aromatic ring structures [35] suggest that the ion accumulation behaviors can be observed for a wide range of ions.

Hydroxylated MWCNTs with outer diameters of 5–50 nm and inner diameters of 2–15 nm were soaked in 0.034 M NaCl solution. This concentration is much lower than that of a normal saline solution (0.15 M) or seawater (~0.25 M). Figure 1A presents a high-angle annular dark field scanning transmission electron microscopy (HAADF-STEM) image clearly showing the formation of an aggregation inside a MWCNT. Further experiments revealed that aggregations frequently formed in the tubes with lengths ranging from 5 to 150 nm (Fig. S2 of PS2 in Supplementary Material (SM)).

Elemental mappings of C, O, Na, and Cl in the aggregation are shown in Fig. 1B. The signal intensities for Na and Cl were as high as that for C. This high concentration was unexpected, as conventional wisdom indicates that the salt concentrations of the solution inside and outside the CNTs should be comparable; thus, only very weak or even negligible Na and Cl signals were expected. Furthermore, the Na and Cl signals were concentrated in the middle of the range for the C signal, and their widths were comparable to the inner diameter of the tube, indicating that the Na and Cl were located inside the CNTs. Thus, our experiments demonstrate the high concentrations of Na and Cl of the aggregation inside the CNTs. In addition, the O signal had a dispersive width comparable to that of the C signal with a higher intensity in the middle. The O signal originates from water in the aggregation and hydroxylation on the tube.

Energy-dispersive X-ray spectroscopy (EDS) was applied to quantify the Na, Cl, and O contents in the aggregations. The Na/O ratios of the aggregations ranged from 0.09 to 1.53 (Fig. 1C), which are 2–3 orders of magnitude higher than that in the salt solution outside the CNT ($6.12 \times 10^{-4}$). The average value of the Na/Cl ratio was $1 \pm 0.3$, suggesting equilibrium of Na and Cl inside the CNTs. We note that the detected O mainly originates from water inside the CNT because the O content in empty tubes was very small (Fig. S3 of PS3 in SM). Furthermore, our analysis indicates that the loss of O inside the CNT during the experiments was not noteworthy (see PS4 in SM [36]). Thus, our observation displays such high salt accumulation of CNTs soaked even in a very dilute solution that the concentration of NaCl inside the CNT reaches two or three orders of magnitude higher than the concentrations of the solution outside the CNT.

The salt accumulation inside the MWCNTs was observed for external solutions with a wide range of concentrations (0.017–0.25 M), which encompasses the concentrations of both normal saline solution and seawater. The EDS results indicate that the Na/O ratios in the aggregations ranged from 0.09 to 5 with average values of 0.5–1.8 (Fig. 1D). The Na/O ratios in the external solution ranged from $3.06 \times 10^{-4}$ to $4.50 \times 10^{-3}$. Thus, these ratios are 2–3 orders of magnitude higher than those of the NaCl solution outside the MWCNTs.

We also examined the Na/O ratios of the aggregations in MWCNTs with different inner diameters (2, 3 and 5 nm) after soaking in 0.034 M NaCl solution. As shown in Fig. S5A of PS5 in SM, upon decreasing the inner diameters from 5 to 2 nm, the Na/O ratios of the aggregations increased (more details are provided in PS5 of SM).

EDS analysis of the salt aggregation inside a MWCNT at the same position was performed six times and revealed Na/O ratios of $0.46 \pm 0.1$, with only a slight fluctuation ($\pm 0.1$) (further details are provided in PS6 of SM [37,38]). Moreover, thermogravimetric analysis (TGA) and Raman radial breathing mode (RBM) measurements were performed under different environmental conditions (with heating, in salt solution, and after drying and vacuum treatment) (see details in PS7 and PS8 of SM [36,39]). These experiments confirmed that the salt aggregations inside CNTs are very stable in these experimental processes and indicate that the high salt accumulation inside CNTs soaked in very dilute salt solutions is robust.

The behaviors of the salt aggregations under high intensity electron beam irradiation further indicate that the aggregations include the aqueous solution. Conventionally, electron-beam irradiation is considered an effective method for heating aqueous phases in MWCNTs [40]. In our experiment, the average electron-beam current densities were fixed in the range of 2.0–2.5 A/cm$^2$. The HAADF-STEM images show that during electron irradiation, the salt aggregation separates into several parts, coalesces, and then separates again (Fig. 2A). The sizes and locations of the separated parts also changed inside the tubes (see Supplementary Video 1 and PS9 in SM). During the initial stage, the salt aggregations mainly exhibited an amorphous structure, although some crystalline-like clusters were detected (Fig. 2B). After prolonged irradiation, NaCl nanocrystals were detected inside the tube (Fig. 2C).

Why do the salt aggregations inside the CNTs have such a high concentration of salt only soaked in a dilute or even very dilute solution? This can be attributed to the hydrated cation–π interactions between the hydrated cations and the π electrons in the aromatic rings of the CNTs. We note that cation–π interactions have been shown to play crucial roles in the structures, dynamic processes, and functions of biological and materials systems [41-47]. However, it was proposed 20 years ago that the cation–π interaction was greatly reduced by the hydrated cations [48,49], and the cation–π interactions between hydrated ions and CNTs in solution have not generally been fully considered. Our previous theoretical studies have shown that this interaction remains strong enough to result in the strong adsorption of hydrated cations on graphitic surfaces (such as CNTs, graphene, graphite, and graphene oxide) because of the polycyclic aromatic ring structure, which includes more π electrons [8,50-53]. By incorporating cation–π interactions with classic all-atoms force fields (see PS1 in SM [54-63] citing Ref. 8), our MD simulations show that the distribution probabilities for Na$^+$ and Cl$^−$ inside a wider CNT with a diameter of 1.4 nm are much higher than the corresponding values in the external solution (Fig. 3A). The average value of the Na/O ratio inside the CNT was 0.33, which is in the range of the corresponding ratio (0.09–1.53) in our experiments. This ratio is over two orders of magnitude higher than the corresponding value of 0.003 in the initial external solution, which is consistent with our experimental observation. Notably, a wave-like structure with peaks and valleys, similar to that of pure water in CNTs, was observed [9]. The inset in Fig. 3A displays the radial distribution, which shows that the water molecules were closer to the CNT surface than the Na$^+$. In the snapshot shown in Fig. 3B, it can be seen that the Na$^+$ around by two water molecules is above one water molecule, indicating that the water molecule is closer to the CNT inner surface. Moreover, we performed theoretical analysis and calculations based on the conventional chemical equilibrium theory to further understand the ion aggregations in wide CNTs (see details in PS10 of SM [64] citing Refs. 8, 52). The Na/O ratio in the aggregation was ~1300 times higher than that in the solution outside the CNT, and the ratio decreased with increasing inner diameter, which is consistent with our experimental observation. We note that, as the dispersion effect [65], the effect of ions on the electronic fluctuations [66] and the electrodynamic confinement effect of ions by nanotubes [67] are not included at DFT method, the van der Waals (vdW) interaction is not well considered in the computation [68,69]. Those effects will further enhance the interaction energy between hydrated cations and CNTs so that it does not influence our conclusion on the high ions enrichment inside CNTs (see more discussion in PS1 of SM).

We think that those effects should be carefully considered in the future in order to get a better quantitative explanation of the experimental observation.

Similar accumulation behavior with high concentration was observed for other chlorine salt solutions (KCl, AuCl$_3$, and PtCl$_4$) (see PS11–13 in SM). Remarkably, even in the solution containing a trace of noble metal salt, the salt accumulation behavior in the CNTs was clearly observed. Fig. 4A presents a HAADF-STEM image of an aggregation inside a tube soaked in a 0.001 M PtCl$_4$ solution. The corresponding distributions of the C, O, Pt, and Cl signals for the elemental mapping are shown in Fig. 4B. The EDS results indicate that the Pt/O ratio in the aggregations was 2–4 orders (mostly 4 orders) of magnitude higher than that of the salt solutions outside the CNTs (Fig. S13 in SM).

We have shown the high salt accumulation of CNTs in dilute and very dilute solutions and even in solutions containing only trace amounts of noble metal salts. We note that in low-concentration salt solutions, salt adsorption has only been reported outside CNT membranes [70]. Inside (8,8)-type CNTs, our previous theoretical study revealed the trapping of a single hydrated ion inside the CNT [8]. Here, unexpectedly, we experimentally observed the formation of high-concentration salt aggregations inside wider CNTs soaked only in dilute and very dilute solutions. This discovery indicates the blocking of fluid flow across the CNTs, which provides an explanation for recent findings [4,23,26] of negligibly low water flow and ion permeation across wide CNTs in solution. These findings will be beneficial for the development of novel ideas related to CNT-based desalination and purification, such as applying an electric field to push the cations away from the inlets of the CNTs as proposed for the thin CNTs in our previous theoretical study [8]. The high salt accumulation could also be used for the collection/extraction of noble metals by using CNTs in solutions containing trace amounts of noble metal salts for the production/recycling of noble metals [34], environmental conservation [71], and directional transport in biological systems [72]. It should be noted that conventionally, noble metal accumulations [33,73,74] in CNTs have only been observed in high-concentration solutions, and a cation-anion pair have been reported within nanotubes [75]. Our computations indicate that this high salt accumulation inside CNTs can be mainly attributed to the strong hydrated cation–π interactions between hydrated cations and π electrons in the aromatic rings of CNTs. Considering the strong cation–π interactions between other hydrated cations (including Li$^+$, Mg$^{2+}$, Ca$^{2+}$, Cu$^{2+}$, Cd$^{2+}$, Cr$^{2+}$, Pb$^{2+}$, Ag$^+$, Pd$^{2+}$, and Rh$^{3+}$) and the aromatic ring structures, similar accumulation behaviors for a wide range of ions can also be expected.

## Acknowledgments:

We thank Prof. Shengli Zhang, Prof. Yusong Tu, Dr. Nan Sheng and Dr. Jige Chen for their constructive suggestions. The supports from NSFC (11675246, 11574339 and 11290164), the National Science Fund for Outstanding Young Scholars (No. 11722548), the Deepcomp7000 and ScGrid of Supercomputing Center, Computer Network Information Center of Chinese Academy of Sciences, the Special Program for Applied Research on Super Computation of the NSFC-Guangdong Joint Fund (the second phase), and the Shanghai Supercomputer Center of China are acknowledged.

**Figure Captions**

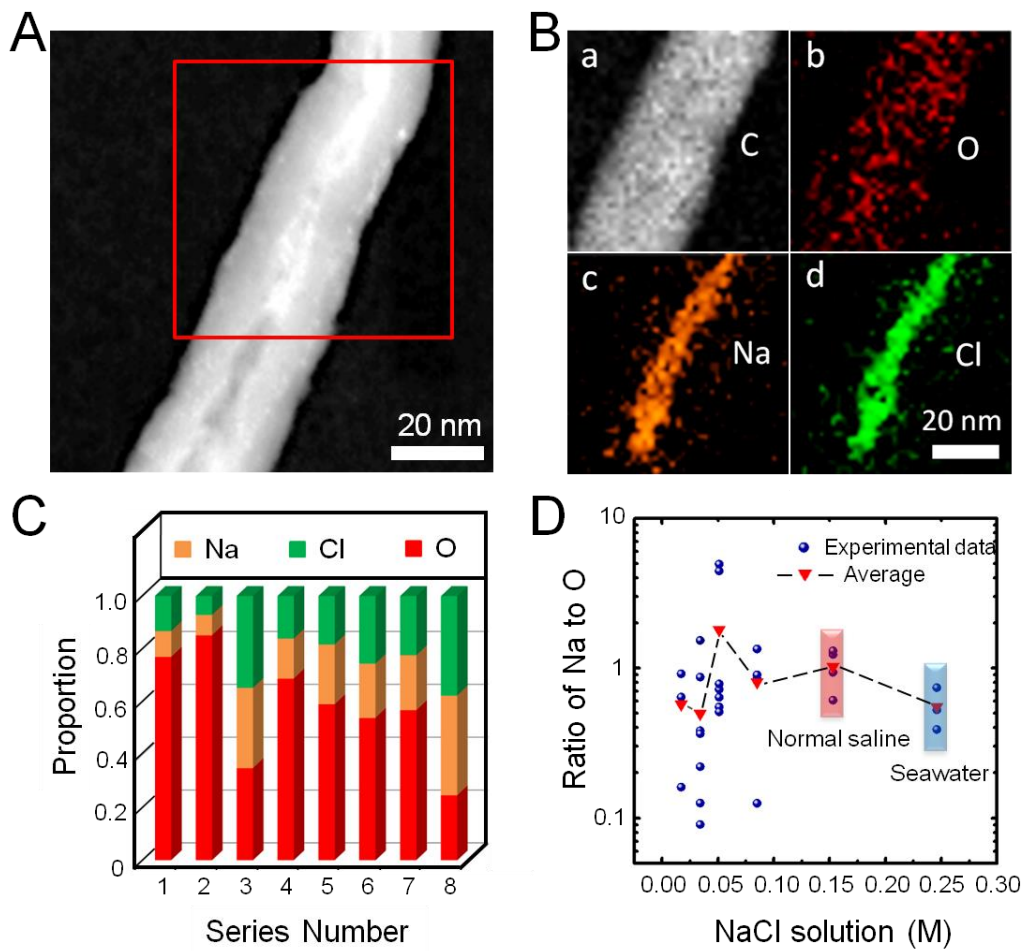

**FIG. 1.** (**A**) HAADF-STEM image of the NaCl aggregation in a MWCNT soaked in 0.034 M NaCl solution. (**B**) C, O, Na, and Cl element mappings of the square frame marked in (**A**) are displayed in (a) - (b), respectively. (**C**) Proportion of Na, Cl, and O elements in salt aggregations inside different MWCNTs for eight series of experiments. (**D**) Na/O ratios in the salt aggregations as a function of the concentration of the soaked NaCl solution outside the MWCNTs. Here, the concentrations corresponding to a normal saline and seawater are marked by pink and cambridge blue rectangles, respectively. The red triangles are the average ratios for several experiments at different concentrations.

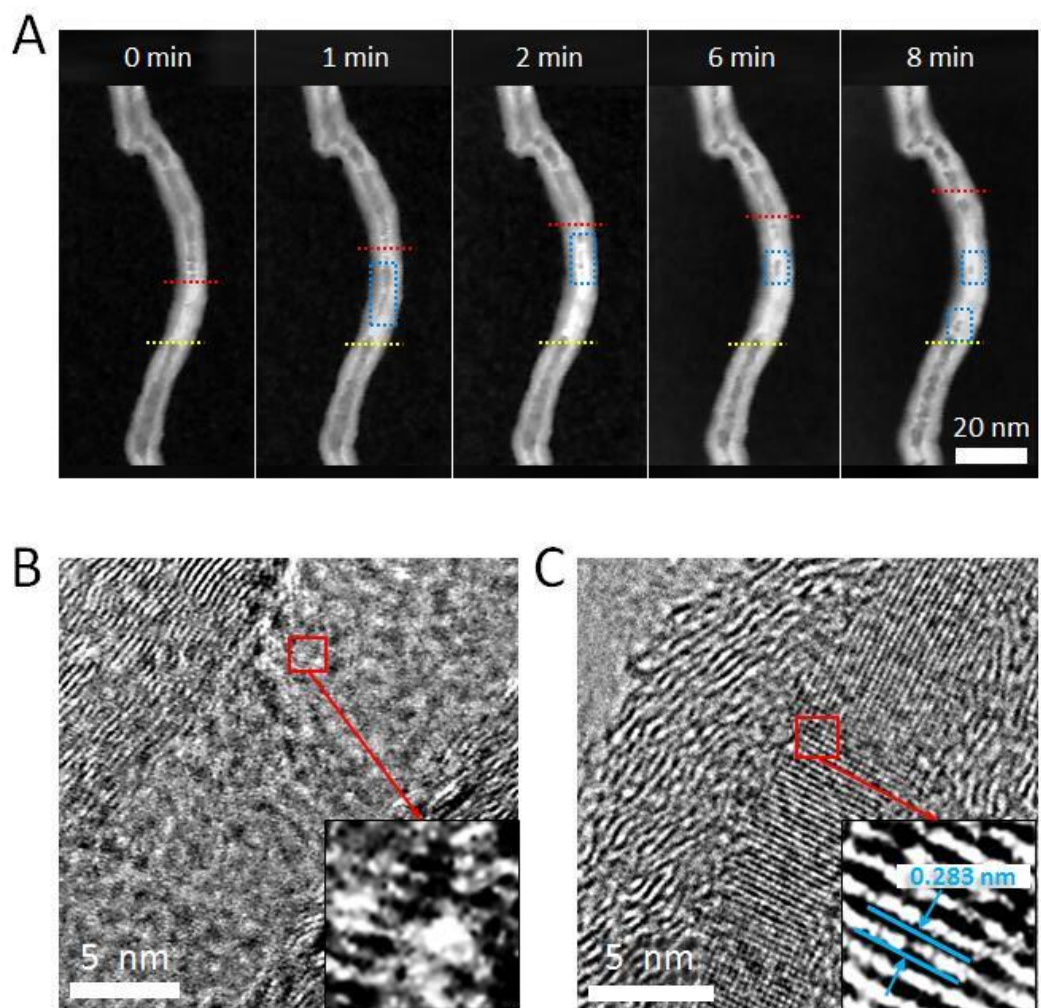

**FIG. 2.** (**A**) HAADF-STEM images of a salt aggregation inside a MWCNT under a high intensity electron beam at different times. Dashed rectangles and lines are used for guidance. (**B** and **C**) TEM images of a segment of MWCNT together with inside salt aggregations in the initial stages and after irradiation for 8 minutes. The electron beam irradiation induced structural changes of the aggregation are displayed by the HRTEM images of insets in (**B**) and (**C**). Here, the interplanar distance of the NaCl crystalline is marked in inset of (**B**).

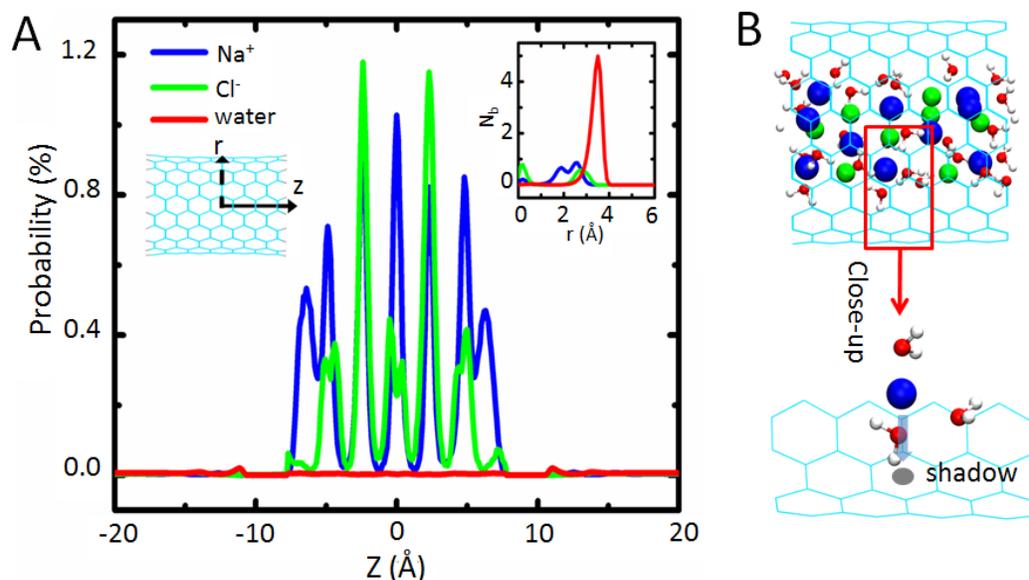

**FIG. 3.** (**A**) Distribution probabilities of $Na^+$, $Cl^-$, and water molecules along the z-direction in the salt solution inside a CNT with a radius of 1.4 nm from MD simulations. Inset is the radial distribution probabilities of $Na^+$, $Cl^-$, and O atoms inside CNT averaged over the perpendicular coordinate. (**B**) Snapshot. Green structures depict the CNT; water molecules and ions are shown with oxygen in red, hydrogen in white, $Na^+$ in blue, and $Cl^-$ in green, respectively.

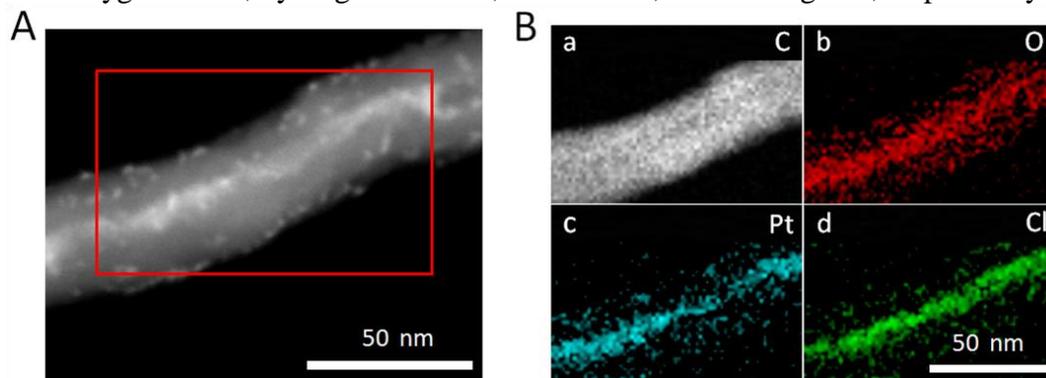

**FIG. 4.** (**A**) HAADF-STEM image of a $PtCl_4$ aggregation inside a MWCNT soaked in a dilute $PtCl_4$ solution of 0.001 M. (**B**) C, O, Pt, and Cl element mappings of the rectangular frame in (**A**) are displayed in (a) - (d), respectively.